\documentclass[reqno,centertags, 12pt]{amsart}
\usepackage{amsmath,amsthm,amscd,amssymb}
\usepackage{latexsym}

\newcommand{\bbR}{{\mathbb{R}}}

\newcommand{\dott}{\,\cdot\,}

\newcommand{\lb}{\label}
\newcommand{\f}{\frac}

\newcommand{\spec}{\text{\rm{spec}}}
\newcommand{\Arg}{\text{\rm{Arg}}}

\newcommand{\bi}{\bibitem}

\newcommand{\beq}{\begin{equation}}
\newcommand{\eeq}{\end{equation}}
\newcommand{\ba}{\begin{align}}
\newcommand{\ea}{\end{align}}
\newcommand{\veps}{\varepsilon}

\allowdisplaybreaks
\numberwithin{equation}{section}

\newtheorem{theorem}{Theorem}[section]
\newtheorem{proposition}[theorem]{Proposition}

\theoremstyle{definition}

\theoremstyle{remark}

\newcommand{\abs}[1]{\lvert#1\rvert}

\begin{document}
\title{Approach to Equilibrium for a Forced Burgers Equation}
\author[W.~Kirsch and B.~Simon]{Werner Kirsch and Barry Simon$^1$}
\date{June 11, 2001}

\footnotetext[1]{Supported in part by NSF Grant No.~DMS-9707661.}
\thanks{$^*$ To appear in Journal of Evolution Equations}

\address{W. Kirsch, Institut f\"ur Mathematik, Ruhr-Universit\"at
Bochum, D-44780 Bochum, Germany}
\email{werner@mathphys.ruhr-uni-bochum.de}

\address{B. Simon,  Division of Physics, Mathematics, and Astronomy,
253-37, California Institute of Technology, Pasadena, CA~91125-3700, USA}
\email{bsimon@caltech.edu}

\begin{abstract} We show that approach to equilibrium in certain forced Burgers
equations is implied by a decay estimate on a suitable intrinsic semigroup
estimate, and we verify this estimate in a variety of cases including a periodic
force.
\end{abstract}
\maketitle

\section{Introduction} \lb{s1}

\bigskip
This paper is a contribution to the literature \cite{Sin1,Sin2,KK} on large 
time asymptotics of the forced Burgers equation
\begin{equation} \lb{1.1}
\f{\partial u_i}{\partial t} + \sum_{j=1}^\nu u_j \f{\partial u_j}{\partial x_i}
= \f12 \, \Delta u_i + \f{\partial V(x)}{\partial x_i},
\qquad t \geq 0, \, x\in\bbR^\nu, \, i=1, \dots, \nu,
\end{equation}
where $u$ is real valued. We will make two assumptions on the initial data
$u_j (x, t=0) =u^{(0)}_j (x)$:
\begin{alignat}{2}
& \text{(i)} \quad && \f{\partial u^{(0)}_i}{\partial x_j} =
\f{\partial u^{(0)}_j}{\partial x_i} \qquad \text{all }i, j \lb{1.2} \\
& \text{(ii)} \quad && \psi_0(x) \equiv \int_0^x \vec u^{(0)} (y)
\cdot dy \in L^\infty. \lb{1.3}
\end{alignat}

\eqref{1.2}, which is vacuous in the standard $\nu=1$ case, implies that the value
of $\psi_0$ given by \eqref{1.3} is independent of the path taken from $0$ to
$x$ in the line integral. Typical of our results is:

\begin{theorem} \lb{T1.1} Let $V$ be a $C^1$ periodic function on $\bbR^\nu$.
Then, there is a unique initial condition $u^{(0)}_\infty (x)$ obeying
\eqref{1.2}, \eqref{1.3} for \eqref{1.1} whose solution is independent of $t$.
Moreover, if $u^{(0)}$ is any other initial data obeying \eqref{1.2}/\eqref{1.3},
then
\begin{equation} \lb{1.4}
\lim_{t\to\infty} \sup_x [ \abs{u(x,t) - u^{(0)}_\infty (x)}] = 0.
\end{equation}
\end{theorem}

\noindent{\it Remark.} \eqref{1.3} does not imply that $u^{(0)}$ is $L^\infty$,
but our proof shows that for $t>0$, $u (\dott, t)\in L^\infty$. Thus, for
$t>0$, the quantity in the limit in \eqref{1.4} is finite.

\smallskip
What is new about our ideas and Theorem~\ref{T1.1} is that there is no regularity
condition on the initial condition $u$ at infinity other than \eqref{1.3}. Previous
approaches require at least that $e^{-\psi_0}$ have some kind of average. To 
understand how we overcome this, we need to begin the proof by reminding the 
reader of the Cole-Hopf transformation. Define
\begin{equation} \lb{1.5}
\varphi_0 (x) = \exp (-\psi_0 (x))
\end{equation}
so that
\begin{equation} \lb{1.6}
u^{(0)}_i = -\varphi^{-1}_0 \, \f{\partial \varphi_0}{\partial x_i}\, .
\end{equation}
$V$ is bounded so
\begin{equation} \lb{1.7}
H = -\tfrac12 \Delta + V
\end{equation}
is bounded below. Thus we add a constant to $V$ so that henceforth
\begin{equation} \lb{1.8}
\inf \spec (H) = 0.
\end{equation}
Define
\begin{equation} \lb{1.9}
\varphi (x,t) = (e^{-tH}\varphi_0)(x).
\end{equation}
Then direct manipulation shows that

\begin{proposition} \lb{P1.2} Let $\psi_0\in L^\infty$ be $C^1$ and let $\varphi_0,
\varphi$ obey \eqref{1.5}, \eqref{1.9}. Then, for $t>0$, $\varphi (x,t) >0$ and 
$\varphi, \nabla\varphi, \Delta\varphi$ are $C^1$ and
\begin{equation} \lb{1.10}
u(x,t) = -\varphi (x,t)^{-1} (\nabla \varphi)(x,t)
\end{equation}
obeys \eqref{1.1} for $t>0$ and $\lim_{t\downarrow 0} u(\dott,t) =
u^{(0)} \equiv \nabla \psi_0$.
\end{proposition}

\noindent{\it Remarks.} 1. It follows from \cite{Simon} that with $V$ a $C^1$ function 
with bounded derivatives that $\varphi$ and $(-\Delta + V)\varphi$ are $C^1$. 
It follows that $\Delta\varphi$ is $C^1$ which by elliptic regularity means 
$\nabla\varphi$ is $C^1$. The Laplacian here and in \eqref{1.1} may be distributional 
rather than classical. If $\nabla V$ is assumed H\"older continuous, we can replace 
these by classical derivatives.

\smallskip
2. As we will discuss below, $e^{-tH}$ maps $L^\infty$ to $L^\infty$ and \eqref{1.9}
is intended in the sense of the $L^\infty$ map.

\smallskip
3. This result does not require that $V$ be periodic; $V$ need only be $C^1$ and in
$L^\infty$. We will use it in this form below.

\medskip
When $V$ is periodic, it has a periodic ground state $\Omega$, that is, a positive
periodic solution of
\begin{equation} \lb{1.11}
(-\tfrac12 \Delta + V) \Omega = 0.
\end{equation}
Thus
\begin{equation} \lb{1.12}
u^{(0)}_\infty (x) = -\Omega (x)^{-1} (\nabla\Omega)(x)
\end{equation}
is a stationary solution of \eqref{1.1}. A natural approach to \eqref{1.4} is to prove that
\begin{equation} \lb{1.13}
\varphi\to c\, \Omega
\end{equation}
and
\begin{equation} \lb{1.14}
\nabla \varphi \to c \nabla\Omega
\end{equation}
both in $L^\infty$. This is essentially what previous works do.

To understand the limitations of this approach and why one can hope to go beyond them,
consider the case $V\equiv 0$. Then $u_\infty =0$, $\Omega\equiv 1$, and
\[
\varphi(x,t) = (2\pi t)^{-\nu/2} \int \exp(-(x - y)^2/2t) \varphi_0 (y)\, dy.
\]
If $(2R)^{-\nu} \int_{\sup_i \abs{y_i} \leq R} \varphi_0 (y)\, d^\nu y \to c$ as $R\to
\infty$, it is not hard to see that $\varphi(x,t) \to c$ as $t\to\infty$ for each fixed
$x$, so \eqref{1.13} holds, but this is not true in general.

For example, if $R_n = e^{e^n}$ and
\[
\varphi_0 (y) = 2 + (-1)^n \quad\text{if}\quad R_n < \sup_i \abs{y_i} < R_{n+1},
\]
then for $t\sim R_n R_{n+1}$, it is not hard to see that $\varphi (0,t) \sim
2 + (-1)^n$ and thus $\varphi(0,t)$ does not have a limit. But in this example, one
can see that $\nabla\varphi$ does go to zero.

Our key observations are that rather than prove \eqref{1.13} and \eqref{1.14}
separately, it suffices to prove that
\begin{equation} \lb{1.15}
\nabla (\varphi/\Omega)\to 0
\end{equation}
and that \eqref{1.15} is equivalent to some estimates on the intrinsic semigroup associated
to $H$. Specifically, let $K_t (x,y)$ be the integral kernel of $e^{-tH}$ and let
\begin{equation} \lb{1.16}
L_t (x,y) = \Omega^{-1} (x) K_t(x,y) \Omega(y)^{-1}.
\end{equation}
$L_t$ is the kernel of a semigroup on $L^2 (\bbR^\nu, \Omega^2 d^\nu x)$.

To prove Theorem~\ref{T1.1}, we will prove two estimates:
\begin{equation} \lb{1.17}
\abs{\partial_x L_t (x,y)} \leq C\, t^{-\nu/2} [\exp (-D(x-y)^2/ t) + \exp (-E\abs{x-y})]
\end{equation}
for suitable $C,D$ and all $t>1$, all $x,y$ and
\begin{equation} \lb{1.18}
\abs{\partial_x L_t (x,y)} \leq C\, t^{-(\nu + 1)/2}
\end{equation}
all $t>1$, all $x,y$.

We will show that if $\Omega$ obeys $0<a\leq\Omega <b$, is $C^3$ and $V\equiv\f12\Omega^{-1}
(\Delta\Omega)$ is bounded and uniformly H\"older continuous, then (even if $V$ is not periodic)
\begin{equation} \lb{1.19}
\abs{\partial_x L_t (x,y)} \leq C\, t^{-\alpha} [\exp (-D(x-y)^2/t)]
\end{equation}
for some $\alpha > \nu/2$, all $t\geq 1$. This will lead to the following generalization
of Theorem~\ref{T1.1}:

\begin{theorem}\lb{T1.3} Let $V$ be a $C^1$ function and suppose that $V$ is bounded and 
uniformly H\"older continuous. Suppose that $-\frac12\Delta + V$ has a ground state $\Omega$ 
obeying $0 < a \leq \Omega \leq b$ for some $a,b$ and all $t$. Then there is a unique initial 
condition $u^{(0)}_\infty (x)$ {\rm(}$=\nabla\Omega/\Omega${\rm)} obeying \eqref{1.2}, 
\eqref{1.3} for \eqref{1.1} whose solution is independent of $t$. Moreover, if $u^{(0)}$ 
is any other initial data obeying \eqref{1.2}/\eqref{1.3}, then
\begin{equation} \lb{1.20}
\lim_{t\to\infty} \sup_x [\abs{u(x,t) - u^{(0)}_\infty (x)}] = 0.
\end{equation}
\end{theorem}

Certain quasiperiodic Schr\"odinger operators have a quasiperiodic ground state 
\cite{Ko}. Thus the above theorem applies to this situation as well; see \cite{Sin2}.

In Section~\ref{s2}, we will reduce Theorems~\ref{T1.1} and \ref{T1.3} to \eqref{1.19}.
In Section~\ref{s3}, we will derive \eqref{1.17} using ideas due to Davies. In
Section~\ref{s4}, we will prove \eqref{1.18} in the periodic case and \eqref{1.19} in
general.

\medskip
We are dedicating this paper to the memory of Tosio Kato, who taught us so much 
about Schr\"odinger operators, about semigroups, and about non-linear equations, 
areas which come together here.

\bigskip
\section{Reduction to Intrinsic Heat Kernel Estimates} \lb{s2}

According to Proposition~\ref{P1.2}, the solution $u$ of \eqref{1.1} is related to
a solution of \eqref{1.9} via \eqref{1.16}. This leads to

\begin{theorem}\lb{T2.1} Suppose there exist $\Omega$ so
\begin{equation} \lb{2.1}
e^{-tH} \Omega = \Omega
\end{equation}
and
\begin{equation} \lb{2.2}
0 < a \leq \Omega(x) \leq b
\end{equation}
and that
\begin{equation} \lb{2.3}
\lim_{t\to\infty} \| \vec\nabla (\varphi (\dott, t)/\Omega (\dott)) \|_\infty =0.
\end{equation}
Then \eqref{1.20} holds with $u^{(0)}_\infty = \nabla\Omega /\Omega$.
\end{theorem}

\begin{proof} By \eqref{1.3}, we have
\[
0 < c_1 \leq \varphi_0 \leq c_2
\]
so by \eqref{2.2}
\[
c_1 b^{-1} \Omega \leq \varphi_0 \leq c_2 a^{-1} \Omega.
\]
Since $e^{-tH}$ is positivity preserving and \eqref{2.1} holds,
\[
c_1 b^{-1} \Omega (x) \leq \varphi(x,t) \leq c_2 a^{-1} \Omega (x).
\]
So by \eqref{2.2} again,
\begin{equation} \lb{2.4}
c_1 b^{-1} a \leq \varphi(x,t) \leq c_2 a^{-1} b.
\end{equation}

Now
\begin{align*}
u(\dott, t) - u^{(0)}_\infty &= (\nabla\varphi)(\dott, t) / \varphi (\dott, t)
-(\nabla\Omega)(\dott)/\Omega(x) \\
&= (\Omega\nabla\varphi - \varphi\nabla\Omega)/ \varphi\Omega \\
&= [\nabla (\varphi (\dott, t)/\Omega)][\Omega/ \varphi].
\end{align*}
Since $\varphi$ and $\Omega$ are uniformly in $t$ and $x$ bounded above and below, we see
that \eqref{1.20} is equivalent to \eqref{2.3}.
\end{proof}

Now consider the unitary map $U:L^2 (\bbR^\nu)\to L(\bbR^\nu, \Omega^2\, dx)$ by $(Uf)(x) =
f(x) \Omega (x)^{-1}$ and let $M$ be the self-adjoint operator $U\! HU^{-1}$ on $L^2 (\bbR^\nu,
\Omega^2\, dx)$. Then, as is well-known (and a direct calculation),
\[
(f, Mf)_{L^2 (\bbR^\nu, \Omega^2\, dx)} = \int (\nabla f)^2 \Omega^2\, dx
\]
or equivalently,
\begin{equation} \lb{2.5}
Mf = -\Delta f - 2(\vec\nabla \Omega) \Omega^{-1} \cdot \vec\nabla f.
\end{equation}

Now let $K_t (x,y)$ be the integral kernel of $e^{-tH}$, that is,
\[
(e^{-tH} f)(x) = \int K_t (x,y) f(y)\, d^\nu y
\]
and let $L_t (x,y)$ be the integral kernel of $e^{-tM}$, that is,
\[
(e^{-tM}f)(x)=\int L_t (x,y) f(y) \Omega^2 (y)\, d^\nu y.
\]
Since $e^{-tM} = Ue^{-tH}U^{-1}$, we see that $L_t$ and $K_t$ are related by \eqref{1.16}.

Now if $\varphi = e^{-tH}\varphi_0$, then
\begin{align*}
[\varphi(\dott, t)/\Omega(\dott)] &= Ue^{-tH}\varphi_0 = e^{-tM} U\varphi_0 \\
&= \int L_t (\dott, y) \varphi_0 (y) \Omega (y)\, d^\nu y.
\end{align*}
Since $\varphi_0$ and $\Omega$ are uniformly bounded, we see that
\[
\abs{\nabla [\varphi(x,t)/\Omega(x)]} \leq c \int \abs{\partial_x L_t (x,y)}\, d^\nu y.
\]
Thus:

\begin{proposition} \lb{P2.2} A sufficient condition for \eqref{2.3} to hold for any
initial $\varphi_0$ {\rm(}coming from a $u^{(0)}_i$ obeying \eqref{1.2}/\eqref{1.3}{\rm)}
is that
\begin{equation} \lb{2.6}
\sup_x \int \abs{\partial_x L_t (x,y)}\, d^\nu y \to 0
\end{equation}
as $t\to\infty$.
\end{proposition}

\begin{theorem} \lb{T2.3} If \eqref{1.19} holds or if \eqref{1.17}/\eqref{1.18}
hold, then \eqref{2.6} holds.
\end{theorem}

\begin{proof} \eqref{1.19} plus scaling implies that
\[
\int \abs{\partial_x L_t (x,y)} \, d^\nu y \leq C_1\, t^{-\alpha} t^{\nu/2}
\]
which goes to zero if $t\to\infty$ since $\alpha > \nu/2$. \eqref{1.17}/\eqref{1.18}
imply 
\[
\abs{\partial_x L_t (x,y)} \leq C \, t^{-\nu/2 - 1/4} [\exp (-D(x-y)^2/2t) +
\exp(-\tfrac12 E\abs{x-y})]
\]
which implies
\[
\int \abs{\partial_x L_t (x,y)}\, d^\nu y \leq C_1 \, t^{-1/4} \pm C_2 \,
t^{-\nu/2 - 1/4}
\]
which goes to zero as $t\to\infty$.
\end{proof}

\bigskip
\section{Exponential-Gaussian Estimates on $\partial_x L_t$} \lb{s3}

Our goal in this section is to explain how one can get \eqref{1.17} from ideas of
Davies \cite{Dav1,Dav2}. His ideas immediately imply an estimate
\begin{equation} \lb{3.1}
\abs{K_t(x,y)} \leq C_\veps \, t^{-\nu/2} \exp (-(x-y)^2/(4+\veps)t)
\end{equation}
for any $\veps >0$. \eqref{3.1} implies (even dropping the Gaussian) that
\[
\| e^{-tH}\|_{L^1\to L^\infty} \leq C\, t^{-\nu/2}
\]
so by interpolation with boundedness on $L^\infty$ (see Simon \cite{Simon} and 
references therein),
\[
\| e^{-tH}\|_{L^1\to L^2} = \| e^{-tH}\|_{L^2 \to L^\infty} \leq C\, t^{-\nu/4}.
\]
Thus
\begin{align*}
\| e^{-(t + is)H}\|_{L^1 \to L^\infty} &\leq \| e^{-tH/2}\|_{L^2 \to L^\infty} \,
\| e^{-isH}\|_{L^2 \to L^2}\,  \| e^{-tH/2}\|_{L^1 \to L^2} \\
&\leq C\, t^{-\nu/2}
\end{align*}
and thus for any $\abs{\theta} < \pi/2$,
\[
\|\exp (-te^{i\theta}H)\|_{L^1 \to L^\infty} \leq C_\theta\, t^{-\nu/2}
\]
which yields
\begin{equation} \lb{3.2}
\abs{K_{t e^{i\theta}} (x,y)}\leq C_\theta\, t^{-\nu/2}.
\end{equation}
By interpolation between \eqref{3.1} and \eqref{3.2}, we see that for complex $t$ in a
section $S_\theta = \{ t\mid \abs{\Arg(t)}\leq\theta\}$ we have if $\theta < \pi/2$,
\begin{equation} \lb{3.3}
\abs{K_t (x,y)} \leq C_\theta \, \abs{t}^{-\nu/2} \exp (-D_\theta (x-y)^2/\abs{t})
\end{equation}
and this implies by a Cauchy estimate that in the same sectors
\[
\abs{\partial_t K_t (x,y)} \leq C_\theta \, \abs{t}^{-\nu/2 -1}
\exp (-D_\theta (x-y)^2/\abs{t})
\]
(where $C_\theta, D_\theta$ can change value from one equation to the next).

Thus for $t\geq 1$ and real,
\begin{equation} \lb{3.4}
\abs{(-\Delta + V)_x K_t (x,y)} \leq C \, \abs{t}^{-\nu/2 -1} \exp (-D (x-y)^2/\abs{t}).
\end{equation}
Since $V$ is bounded, \eqref{3.3} and \eqref{3.4} imply that for $t\geq 1$,
\[
\abs{(-\Delta_x + 1) K_t (x,y)} \leq C \, \abs{t}^{-\nu/2} \exp (-D(x-y)^2 /\abs{t}).
\]
But $\partial_x (-\Delta_x + 1)^{-1}$ has an explicit convolution integral kernel which is
$L^1$ at short distances and exponentially decaying at large. This implies for $t\geq 1$
\[
\abs{(\partial_x K_t) (x,y)} \leq C\, t^{-\nu/2} [\exp (-D(x-y)^2/t) +
\exp(-E\abs{x-y})]
\]
which, with \eqref{3.3} and the formula \eqref{1.16}, implies \eqref{1.17}. We summarize:

\begin{theorem} \lb{T3.1} The estimate \eqref{1.17} holds for any potential $V$ obeying
the hypothesis of Theorem~\ref{T1.3}.
\end{theorem}

\bigskip
\section{Improved Time Decay Estimates on $\partial_t L_t$} \lb{s4}

Here the goal is to show that
\begin{equation} \lb{4.1}
\abs{(\partial_x L_t) (x,y)} \leq C\, t^{-\alpha}
\end{equation}
for some $\alpha > \nu/2$. We believe the estimate holds with $\alpha = \nu/2 + 1/2$
and have proven this if $V$ is periodic: One makes a Bloch wave decomposition 
\cite{RS} to write the semigroup as an integral over the Brillouin zone and a 
Gaussian approximation to control the resulting integral. One uses the fact that 
the minimum of the bottom band is known to be a unique point with a strictly 
quadratic minimum \cite{KS}. In general, we rely on estimates of Porper-Eidel'man 
\cite{PE1} and only get $\alpha > \nu/2$. (In a later paper \cite{PE2}, they get 
$\alpha = \nu/2 + 1/2$ for the case $p=1$ below that does not accommodate our situation.)

Indeed, \eqref{4.1} is exactly Corollary~3.4 of their paper which they prove for
fundamental solutions of equations of the form
\begin{equation} \lb{4.2}
p(x) \partial_t u = \nabla \cdot (a (t,x)\nabla u),
\end{equation}
where $p,a$ and $\nabla a$ are all H\"older continuous.

But \eqref{2.5} can be rewritten
\[
Mf = -\Omega^{-2} \nabla \cdot (\Omega^2 \nabla f)
\]
so
\[
\partial_t u = -Mu
\]
is of the form \eqref{4.2} where
\[
p = \Omega^2 \quad\text{and}\quad a_{ij} = \Omega^2 \delta_{ij}.
\]Thus, their result applies so long as $\Omega$ is $C^1$ with $\nabla\Omega$ uniformly
H\"older continuous.

Since $\Delta\Omega = V\Omega$ with $V\in C^1$ and $\Omega$ a priori bounded, we see that
\[
\nabla\Omega = \nabla (-\Delta + 1)^{-1} (-V+1) \Omega.
\]
Since $(-V+1)\Omega$ is bounded, the explicit integral kernel for $\nabla (-\Delta + 1)^{-1}$
shows $\nabla\Omega$ is uniformly H\"older continuous of any order less than $1$. Thus, their
Corollary~3.4 applies and \eqref{4.1} holds.

\bigskip
\noindent{\bf Acknowledgments.}  W.K.~would like to thank T.~Tombrello for
the hospitality of Caltech where this research was done. We would like to thank
E.B.~Davies, T.~Coulhon, A.~Grigoryan, Y.~Pinchover, and D.~Robinson for valuable
correspondence.

\bigskip


\end{document}